\documentclass[pra,aps,showpacs,twocolumn,superscriptaddress]{revtex4}

\usepackage{amsmath}
\usepackage{amsfonts}
\usepackage{bm}
\usepackage{graphicx}

\begin{document}

\title{Collision dynamics of Skyrmions in a two-component Bose--Einstein
condensate}

\author{Tomoya Kaneda}
\affiliation{Department of Engineering Science, University of
Electro-Communications, Tokyo 182-8585, Japan}

\author{Hiroki Saito}
\affiliation{Department of Engineering Science, University of
Electro-Communications, Tokyo 182-8585, Japan}

\date{\today}

\begin{abstract}
The dynamics of Skyrmions in a two-component Bose-Einstein condensate are
numerically investigated in the mean-field theory.
When two Skyrmions collide with each other, they are first united and
then scattered into various states.
For head-on collisions, Skyrmions with unit winding number are scattered.
The collision dynamics with an impact parameter are shown to depend on the
relative phase.
These dynamic processes are characterized by integer winding numbers.
\end{abstract}


\maketitle

\section{Introduction}

A quantized vortex in a superfluid is a topological excitation that
reflects the U(1) manifold of the order-parameter space.
For multicomponent superfluids with larger degrees of freedom in the
order-parameter space, there exist a rich variety of topological
excitations, such as spin vortices, monopoles, and
Skyrmions~\cite{Skyrme}, which have been realized in a superfluid
$^3{\rm He}$~\cite{Blaau}, Bose-Einstein condensates (BECs) of ultracold
gases~\cite{Leslie,Choi,Ray14,Ray15}, and
exciton-polariton superfluids~\cite{Lagou,Hivet}.

When two or more topological excitations are generated in proximity to
each other, they can exhibit interesting dynamics.
The simplest example is a quantized vortex-antivortex pair (called a
vortex dipole), which travels at a constant velocity in a uniform system. 
Such a topological object in a BEC has been studied
theoretically~\cite{Frisch,Aioi} and realized
experimentally~\cite{Inouye,Neely,Freilich,Kwon}.
A pair of quantized vortices with the same circulation rotate around one
another~\cite{SasakiL,Navarro}.
When two quantized vortex lines approach each other, they interact with
one another~\cite{Serafini}, and a reconnection
occurs~\cite{Schwarz,Koplik93}.
Two copropagating quantized vortex rings show leapfrogging
dynamics~\cite{Sasaki11,Caplan}, such as those in classical fluids, and
when they collide with each other, they merge and split again into vortex
rings~\cite{Koplik96,Caplan}.
For multicomponent or spinor BECs, interaction between half-quantum
vortices~\cite{Eto}, reconnection of $1/3$-vortices~\cite{Kobayashi},
the dynamics of spin-vortex dipoles~\cite{Kaneda}, and generation of
multiple Skyrmions~\cite{Sasaki} have been predicted.
Recently, the collision of half-quantum vortices in a spin-1 BEC was
observed~\cite{Seo}.

In the present paper, we investigate the collision and scattering dynamics
of Skyrmions in a two-component BEC.
Although the scattering of Skyrmions has been studied in the context of
high energy physics~\cite{Allder,Battye97,Foster}, previous studies on
Skyrmions in a two-component BEC have mainly focused on their static
properties~\cite{Ruo,Khawaja,Battye,Savage,Ruo04,Wuster,Herbut}.
A Skyrmion in a two-component BEC consists of a quantized vortex ring in
one component, whose core is occupied by a quantized vortex of the other
component.
A Skyrmion therefore travels at a constant velocity, since a vortex ring
has a momentum along the symmetry axis.
Let us consider a situation in which two Skyrmions move toward and collide
with each other.
The topology of a Skyrmion is characterized by an integer winding number,
and the sum of the winding numbers of the two Skyrmions, $W_1 + W_2$, is
conserved during their collision, if the wave functions are always
restricted to the SU(2) manifold.
When the winding number is conserved, we find that $W_1 + W_2$ Skyrmions
with unit winding number are scattered after the collision.
We present the collision dynamics of Skyrmions with various winding
numbers.
For off-axis collisions with finite impact parameters, we show that the
dynamics depend on the relative phases between the two Skyrmions, whereas
the head-on collisions are independent of the relative phases.

This paper is organized as follows.
Section~\ref{s:formulation} introduces a Skyrmion in a two-component BEC
and discusses the problem that we will consider in this paper.
Section~\ref{s:result} shows numerical results for the dynamics of
Skyrmions.
Section~\ref{s:conc} presents our conclusions for this study.

\section{Formulation of the problem}
\label{s:formulation}

\subsection{Skyrmion in a two-component BEC}

First, we briefly review a Skyrmion in a two-component BEC.
The mean-field energy of a two-component BEC in a three-dimensional (3D)
free space is given by
\begin{equation} \label{E}
E =  \int d\bm{r} \left( -\sum_{j=1}^2 \psi_j^* \frac{\hbar^2}{2m_j}
\nabla^2 \psi_j + \sum_{j, j'} \frac{g_{jj'}}{2} |\psi_j|^2
|\psi_{j'}|^2 \right),
\end{equation}
where $\psi_j$ and $m_j$ are the macroscopic wave function and the mass of
the atom of the $j$th component, respectively.
The interaction coefficient in Eq.~(\ref{E}) is defined as
$g_{jj'} = 2\pi\hbar^2 a_{jj'} (m_j^{-1} + m_{j'}^{-1})$, where $a_{jj'} =
a_{j'j}$ is the $s$-wave scattering length between the atoms in components
$j$ and $j'$.

When $g_{11} \simeq g_{22} \simeq g_{12}$ and the interaction energy is
much larger than the kinetic energy, the total density $\rho = |\psi_1|^2
+ |\psi_2|^2$ is approximately uniform.
The wave functions are then written as
\begin{equation} \label{state}
\bm{\Psi}(\bm{r}) = \left( \begin{array}{c} \psi_1(\bm{r}) \\
\psi_2(\bm{r}) \end{array} \right) = \sqrt{\rho} \left( \begin{array}{c}
\xi_1(\bm{r}) \\
\xi_2(\bm{r}) \end{array} \right),
\end{equation}
where $|\xi_1|^2 + |\xi_2|^2 = 1$.
The two-component state is thus described by the SU(2) manifold.
A Skyrmion is defined as a topological state in which $\Psi(\bm{r})$
goes to the same state $\Psi_0$ at infinity, i.e.,
$\lim_{r \rightarrow \infty} \Psi(\bm{r}) = \Psi_0$.
Such a state can be described by  a map of SU(2) on the sphere $S^3$ in 4D
space.
(This can be understood from a lower-dimensional analogy, e.g., a map of
SU(2) on the sphere $S^2$ in 3D space: imagine that the sphere is cut open
from a point on the sphere, and the cut edge is expanded to infinity; this
is equivalent to a 2D plane satisfying $\lim_{r \rightarrow \infty}
\Psi(\bm{r}) = \Psi_0$.)
Topologically, the way in which SU(2) is mapped onto $S^3$ is expressed by
$\pi_3(SU(2)) = \mathbb{Z}$, which indicates that the Skyrmion state is
characterized by an integer topological number.

Separating the complex variables $\xi_j$ in Eq.~(\ref{state}) into their
real and imaginary parts, $\xi_j = c_j + i d_j$, we have $|\xi_1|^2 +
|\xi_2|^2 = c_1^2 + d_1^2 + c_2^2 + d_2^2 = 1$, which is a unit sphere in
4D.
Using polar coordinates as $c_1 = \sin\alpha \sin\beta \sin\gamma$,
$d_1 = \sin\alpha \sin\beta \cos\gamma$, $d_2 = \sin\alpha \cos\beta$, and
$c_2 = \cos\alpha$, Eq.~(\ref{state}) becomes
\begin{equation}
\bm{\Psi}(\bm{r}) = \sqrt{\rho} \left( \begin{array}{c} i
\sin\alpha(\bm{r}) \sin\beta(\bm{r}) e^{-i\gamma(\bm{r})} \\
\cos\alpha(\bm{r}) + i \sin\alpha(\bm{r}) \cos\beta(\bm{r}) \end{array}
\right).
\end{equation}
Noting that the area of the unit sphere in 4D is $\int_0^{\pi} d\alpha
\int_0^{\pi} d\beta \int_0^{2\pi} d\gamma \sin^2\alpha \sin\beta =
2\pi^2$, the number of times that $S^3$ (3D space) covers SU(2)
(two-component state) is expressed as
\begin{equation} \label{W}
W = \frac{1}{2\pi^2} \int d\bm{r} \sin^2\alpha(\bm{r}) \sin\beta(\bm{r})
{\rm det} \left( \frac{\partial(\alpha, \beta, \gamma)}{\partial(x, y, z)}
\right),
\end{equation}
where ${\rm det} (\cdots)$ is the Jacobian.
The winding number $W$ is an integer reflecting $\pi_3(SU(2)) =
\mathbb{Z}$.
In the numerical analysis, it is convenient to express the two-component
state as
\begin{equation}
\bm{\Psi}(\bm{r}) = \sqrt{\rho} \left( \begin{array}{c} 
\cos \frac{\theta(\bm{r})}{2} e^{i \phi_1(\bm{r})} \\
\sin \frac{\theta(\bm{r})}{2} e^{i \phi_2(\bm{r})}
\end{array} \right).
\end{equation}
Since the area of the unit sphere $c_1^2 + d_1^2 + c_2^2 + d_2^2 = 1$ is
written as $\int_0^\pi d\theta \int_0^{2\pi} d\phi_1 \int_0^{2\pi} d\phi_2
\frac{1}{4} \sin\theta = 2\pi^2$, the winding number $W$ is given by
\begin{equation} \label{W2}
W = \frac{1}{8\pi^2} \int d\bm{r} \sin\theta(\bm{r}) {\rm det} \left(
\frac{\partial(\theta, \phi_1, \phi_2)}{\partial(x, y, z)} \right),
\end{equation}
which is also an integer.

\subsection{Dynamics of the system}

In the mean-field approximation, the dynamics of the two-component BEC is
described by the Gross-Pitaevskii (GP) equation,
\begin{subequations} \label{GP}
\begin{eqnarray}
i \hbar \frac{\partial \psi_1}{\partial t} & = & -\frac{\hbar^2}{2m_1}
\nabla^2 \psi_1 + g_{11} |\psi_1|^2 \psi_1 + g_{12} |\psi_2|^2 \psi_1, \\
i \hbar \frac{\partial \psi_2}{\partial t} & = & -\frac{\hbar^2}{2m_2}
\nabla^2 \psi_2 + g_{22} |\psi_2|^2 \psi_2 + g_{12} |\psi_1|^2 \psi_2.
\end{eqnarray}
\end{subequations}
In the following, we assume $m_1 = m_2 \equiv m$ and $g_{11} = g_{22} =
g_{12} \equiv g$.
The characteristic length and velocity of the system are given by the
healing length $\xi = \hbar / (m g n_0)^{1/2}$ and sound velocity $v_s
= (g n_0 / m)^{1/2}$, where $n_0$ is the uniform density far from the
Skyrmions.
The characteristic time is given by $\tau = \xi / v_s = \hbar / (g n_0)$.

We numerically solve the GP equation using the pseudospectral method.
The initial state of a Skyrmion is prepared as follows.
First we numerically imprint a Skyrmion as
\begin{subequations} \label{psiini}
\begin{eqnarray}
\psi_1(\bm{r}) & = & \sqrt{n_0} e^{-\left[ \left(\sqrt{\Delta x^2
+ \Delta y^2} - R_s \right)^2 + \Delta z^2 \right] / r_s^2} 
e^{i n \phi}, \\
\psi_2(\bm{r}) & = & \sqrt{n_0 - |\psi_1(\bm{r})|^2} e^{i \ell \chi},
\end{eqnarray}
\end{subequations}
where $\Delta \bm{r} = \bm{r} - \bm{r}_0$, and $\bm{r}_0$ is the center of
the Skyrmion, $R_s$ and $r_s$ are constants that determine the shape of
the Skyrmion, $n$ and $\ell$ are integers, $\phi = {\rm arg} (\Delta x + i
\Delta y)$, and $\chi = {\rm arg}(\sqrt{\Delta x^2 + \Delta y^2} - R_s + i
\Delta z) - {\rm arg}(\sqrt{\Delta x^2 + \Delta y^2} + R_s + i \Delta
z)$.
The winding number of the state in Eq.~(\ref{psiini}) is $W = n \ell$.
We then perform the imaginary-time propagation for some time (typically
$20 \tau$), in which $i$ on the left-hand side of Eq.~(\ref{GP}) is
replaced by $-1$, and the wave functions are normalized to their initial
values in every time step.
By the imaginary-time propagation, the excess energy of the Skyrmion
imprinted by Eq.~(\ref{psiini}) is relaxed.
Starting from the wave functions obtained by the imaginary-time
propagation, we perform the real-time propagation to study the dynamics of
the system, where a small initial noise is added to the initial state to
break the symmetry.
The size of the numerical mesh is typically $(256)^3$ or $(512)^3$.
The periodic boundary condition is imposed by the pseudospectral method,
which does not affect the dynamics of Skyrmions located near the center of
the numerical space.

\section{Numerical Results}
\label{s:result}

\subsection{Dynamics of a single Skyrmion}

\begin{figure}[tbp]
\includegraphics[width=8cm]{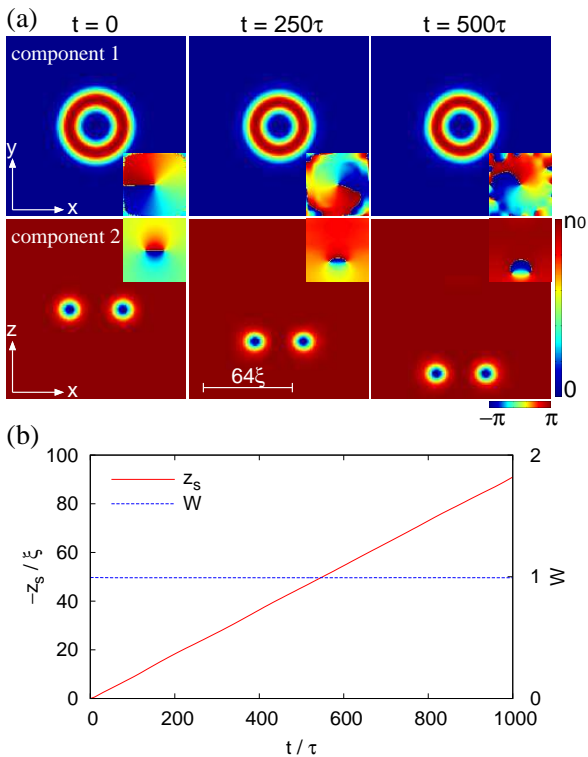}
\caption{
(Color online) Dynamics of a single Skyrmion.
(a) Cross-sectional density profiles of component 1 at $z = z_s$ (upper
panels) and those of component 2 at $y = 0$ (lower panels), where $z_s$ is
the $z$ coordinate of the position of the vortex-ring core in component
2.
The insets show the phase profiles.
The field of view is $128\xi \times 128\xi$.
(b) Time evolution of $z_s$ (red solid line) and the winding number $W$
(blue dashed line).
}
\label{f:single}
\end{figure}
First we numerically investigate the dynamics of a single Skyrmion.
The initial state is given by Eq.~(\ref{psiini}) with $R_s = 20 \xi$,
$r_s = 10 \xi$, and $n = \ell = 1$.
Component 1 has a donut shape with a quantized circulation, which is held
by the quantized vortex ring in component 2, as shown in
Fig.~\ref{f:single}(a).
Since the vortex ring in component 2 has a momentum, the Skyrmion travels
in the $-z$ direction.
We find that the shape of the Skyrmion remains almost unchanged for a long
time~\cite{Note}.

Figure~\ref{f:single}(b) shows the time evolution of the position $z_s$ of
the Skyrmion and its winding number $W$.
The position $z_s$ is defined as the $z$ coordinate of the core of the
vortex ring in component 2.
From the slope of $z_s$ in Fig.~\ref{f:single}(b), the Skyrmion is found
to travel at a constant velocity $\simeq 0.09 v_s$.
For a single-component superfluid, a vortex ring travels at a velocity
$v_r \simeq \frac{\hbar}{2 R_r m} \log \frac{8 R_r}{r_r}$, where $R_r$ is
the radius of a ring and $r_r$ is the size of the vortex
core~\cite{Fetter}.
Substituting $R_s$ and $r_s$ into $R_r$ and $r_r$, we obtain $v_r \simeq
0.07 v_s$, which is in reasonable agreement with the Skyrmion velocity in
Fig.~\ref{f:single}.
The winding number $W$ is calculated using Eq.~(\ref{W2}), which is always
$\simeq 1$ during the time evolution.
The deviation from 1 is less than 1 \%, which is due to the numerical
error from the spatial discretization.

\subsection{Collision dynamics of two Skyrmions}

\begin{figure}[tbp]
\includegraphics[width=8cm]{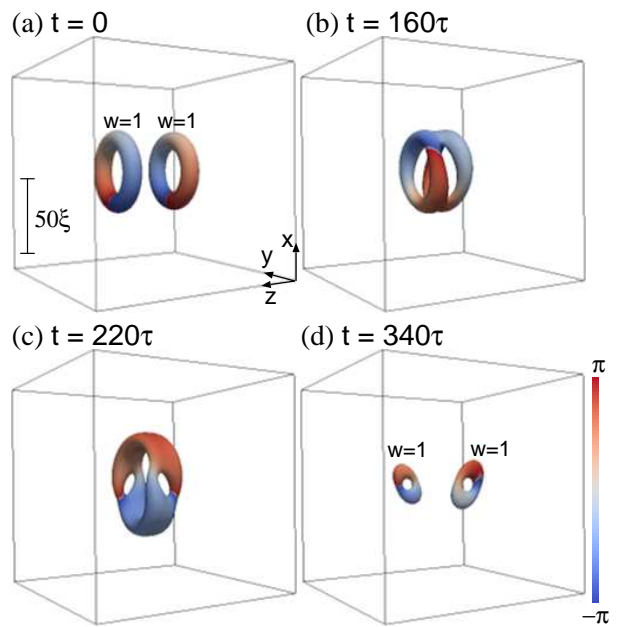}
\caption{
(Color online) Dynamics of a collision and scattering of two Skyrmions.
Isodensity surfaces of component 1 at $|\psi_1|^2 = 0.5 n_0$ are shown;
the color represents the phase at the surface.
Each Skyrmion in (a) has a winding number $W = 1$, and the total winding
number is $W_{\rm tot} = 2$.
The parameters for preparing the initial Skyrmions are $R_s = 20 \xi$,
$r_s = 10 \xi$, ${\bm{r}_0} = (0, 0, \pm 20 \xi)$, $n = \pm 1$, and $\ell
= \pm 1$.
The size of each box is $(128 \xi)^3$.
See the Supplemental Material for a movie of the dynamics~\cite{SM}.
}
\label{f:one_one}
\end{figure}
We first examine the dynamics of the head-on collision of two Skyrmions,
where each winding number is $W = 1$, and the total winding number is
$W_{\rm tot} = 1 + 1 = 2$.
Figure~\ref{f:one_one}(a) is the initial state, with the isodensity
surfaces of component 1 shown.
The Skyrmions travel toward each other, and a head-on collision occurs at
$t \simeq 160\tau$; at this time, the two rings of component 1 touch at
two regions [Fig.~\ref{f:one_one}(b)].
When the two rings merge with each other, two quantized vortices are
created on the ring [Fig.~\ref{f:one_one}(c)].
The ring is then divided into pieces, and the two quantized vortices on
the ring become independent rings, which then become two Skyrmions with
unit winding number [Fig.~\ref{f:one_one}(d)].
After that, the two small Skyrmions are scattered in directions
perpendicular to the incident directions.
The part of component 1 that is not contained in the scattered Skyrmions
spreads into the surrounding component 2 without topological structure,
which has disappeared from Fig.~\ref{f:one_one}(d).
In this dynamics, the deviation of the total winding number $W_{\rm tot}$
from the initial value of 2 is less than 1\%.
The variation in the winding number is due to the roughness of the
numerical mesh.

\begin{figure}[tbp]
\includegraphics[width=8cm]{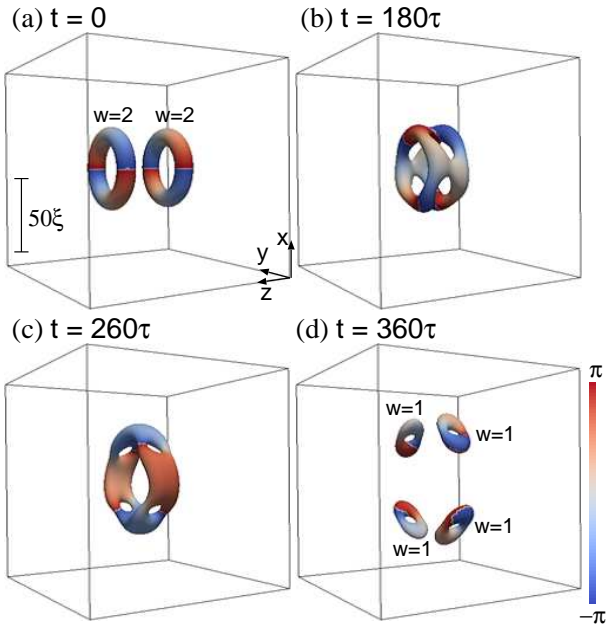}
\caption{
(Color online) Isodensity surfaces of component 1; the color represents
the phase at the surface. 
Each Skyrmion in (a) has winding number 2 ($n = \pm 2$ and $\ell = \pm
1$), and the total winding number is $W_{\rm tot} = 4$.
Other conditions are the same as those in Fig.~\ref{f:one_one}.
See the Supplemental Material for a movie of the dynamics~\cite{SM}.
}
\label{f:two_two}
\end{figure}
Figure~\ref{f:two_two} shows the dynamics of a head-on collision of two
Skyrmions with $W = 2$ ($W_{\rm tot} = 2 + 2 = 4$).
The two donuts of component 1 merge to form a multiply-connected shape
containing four quantized vortices, as shown in Fig.~\ref{f:two_two}(c).
The four small Skyrmions with $W = 1$ are then scattered to the $x$-$y$
directions.
In this dynamics, the deviation of $W_{\rm tot}$ from 4 is less than 1
\%.
After the scattering, we only observe Skyrmions with $W = 1$;
Skyrmions with $W \geq 2$ are never produced.
This is probably because the energy of $n$ Skyrmions with $W = 1$ is less
than that of a single Skyrmion with $W = n$.

\begin{figure}[tbp]
\includegraphics[width=8cm]{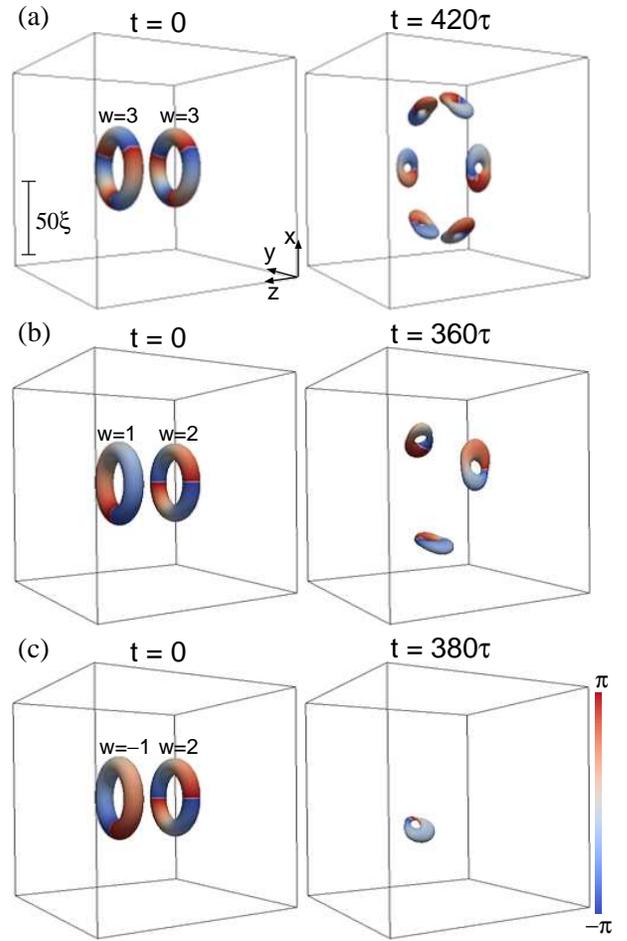}
\caption{
(Color online) Isodensity surfaces of component 1; the color represents
the phase at the surface. 
The initial states of two Skyrmions and the states after their collision
are shown.
(a) Winding numbers are 3 and 3 ($W_{\rm tot} = 6$).
(b) Winding numbers are 1 and 2 ($W_{\rm tot} = 3$).
(c) Winding numbers are $-1$ and 2 ($W_{\rm tot} = 1$).
Other conditions are the same as those in Fig.~\ref{f:one_one}.
All the Skyrmions in the right-hand panels have unit winding number.
See the Supplemental Material for a movie of the dynamics~\cite{SM}.
}
\label{f:other}
\end{figure}
Figure~\ref{f:other} shows the collision of two Skyrmions with various
winding numbers.
In Fig.~\ref{f:other}(a), two Skyrmions with $W = 3$ ($W_{\rm tot} = 3 + 3
= 6$) collide with each other, and six Skyrmions with unit winding number
are scattered after the collision, which is similar to the dynamics in
Figs.~\ref{f:one_one} and \ref{f:two_two}.
Figure~\ref{f:other}(b) shows the case of Skyrmions with different winding
numbers, $W = 1$ and $W = 2$ ($W_{\rm tot} = 3$).
After the collision, the three Skyrmions with unit winding number are
scattered.
Figure~\ref{f:other}(c) shows the case with $W = -1$ and $W = 2$.
Since the total winding number is $W_{\rm tot} = 1$, only a single small 
Skyrmion with unit winding number is left after the collision, where most
of component 1 has spread into component 2 without topological structure.
We also found that a head-on collision between two Skyrmions with
$W_{\rm tot} = 0$ (for example, $W = -1$ and $W = 1$) generates no
topological structure, and component 1 spreads into component 2 after the
collision (data not shown).

\begin{figure}[tbp]
\includegraphics[width=8cm]{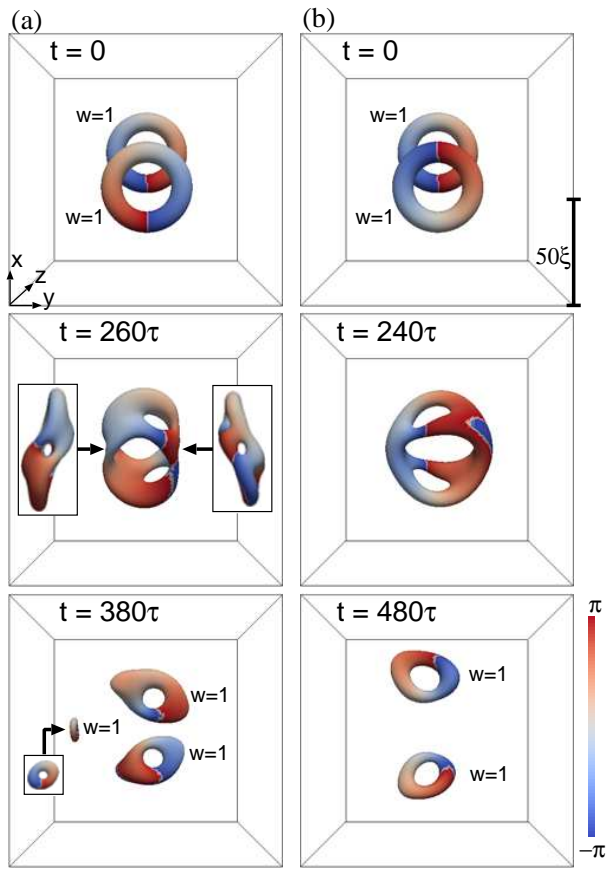}
\caption{
(Color online) Dynamics of off-axis collisions of Skyrmions with $W = 1$
($W_{\rm tot} = 2$) and $\bm{r}_0 = (\pm 10 \xi, 0, \pm 20 \xi)$ (the
impact parameter is $20 \xi$).
Isodensity surfaces of component 1 are shown; the color represents the
phase at the surface.
The initial phase differences between the two donut-shaped components are
0 in (a) and $\pi$ in (b).
Other conditions are the same as those in Fig.~\ref{f:one_one}.
See the Supplemental Material for a movie of the dynamics~\cite{SM}.
}
\label{f:offaxis}
\end{figure}
Figure~\ref{f:offaxis} shows the dynamics of off-axis collisions of
Skyrmions with $W = 1$, where the impact parameter is $20 \xi$.
The difference between Figs.~\ref{f:offaxis}(a) and \ref{f:offaxis}(b) is
only the initial phase of component 1: component 1 contained in the one
Skyrmion is multiplied by $e^{i \eta}$ with $\eta = \pi$ in (b).
When the two donut shapes touch each other in Fig.~\ref{f:offaxis}(a),
four quantized vortices are created between them ($t = 260 \tau$).
One of these decreases the winding number by one, and the remaining three
vortices increase the winding number by three; thus, the total winding
number remains $W_{\rm tot} = 3 - 1 = 2$.
As they split into Skyrmions with $W = \pm 1$, the one with $W = -1$
(corresponding to the vortex in the right-hand inset) disappears; this
increases the total winding number by one, giving $W_{\rm tot} = 3$ ($t =
380 \tau$).
The non-conservation of the total winding number indicates that the
total density vanishes at some point, at which the wave function deviates
from the SU(2) manifold.
In Fig.~\ref{f:offaxis}(b), one of the initial donuts of component 1 is
multiplied by $e^{i \eta} = e^{i\pi}$.
In this case, two quantized vortices are created when the two donuts
unite ($t = 240 \tau$), and they are scattered as two Skyrmions with $W =
1$ ($t = 480 \tau$).
The total winding number $W_{\rm tot} = 2$ is conserved for the dynamics
in Fig.~\ref{f:offaxis}(b).

Thus, the collision dynamics of Skyrmions with a finite impact parameter
depend on the initial relative phase $\eta$ between the two donuts of
component 1.
We also found that the same is true for the oblique collisions.
However, for head-on collisions as in Figs.~\ref{f:one_one}-\ref{f:other},
the initial relative phase $\eta$ is not important, since it is described
by $e^{i L_z \eta / W_{\rm tot}} \Psi$, where $L_z$ is the $z$ component
of the angular momentum operator.
Therefore, when $\eta \neq 0$, the entire dynamics is rotated around the
$z$ axis.

\section{Conclusions}
\label{s:conc}

We have numerically investigated the dynamics of collision and scattering
of Skyrmions in a two-component BEC.
A Skyrmion in a two-component BEC is composed of a quantized vortex ring
in one component, whose donut-shaped core is occupied by a quantized
vortex of the other component.
Since a vortex ring has a momentum, a Skyrmion travels at a constant
velocity, as shown in Fig.~\ref{f:single}.
When two Skyrmions are prepared and collide with each other, various
dynamics can be observed; these depend on the winding numbers, impact
parameters, collision angles, and phase differences.
When two Skyrmions, which both have $W = 1$, collide, they first merge
into a Skyrmion with $W_{\rm tot} = 2$, and then separate into two
Skyrmions with $W = 1$ and 1 and are scattered (Fig.~\ref{f:one_one}).
The scattering of Skyrmions with unit winding number occurs even for a
collision with $W = 2$ and 2.
In this case, four Skyrmions with unit winding number are scattered after
the collision (Fig.~\ref{f:two_two}).
Similar dynamics are also observed for $W_{\rm tot} = 3 + 3$, $W_{\rm tot}
= 1 + 2$, and $W_{\rm tot} = -1 + 2$ collisions, where, respectively, 6,
3, and 1 Skyrmion(s) with unit winding number are scattered
(Fig.~\ref{f:other}).
These dynamics imply that $W_{\rm tot}$ Skyrmions with unit winding number
are always scattered following head-on collisions in which the total
winding number is conserved.
For off-axis collisions, the dynamics depend on the relative phase of the
two Skyrmions (Fig.~\ref{f:offaxis}).

The dynamics of Skyrmions studied in this paper can be reproduced
experimentally, if Skyrmions are created in a sufficiently large BEC in a
controlled manner.
Dynamical creation of Skyrmions in a two-component BEC has been proposed
in Refs.~\cite{Sasaki,Nitta}.
Once two Skyrmions are created, their trajectories may be controllable by
external potentials~\cite{Aioi}, which would enable us to collide
Skyrmions at the desired angle and with the desired impact parameter.

\begin{acknowledgments}
We thank Y. Kawaguchi for fruitful discussion.
This work was supported by JSPS KAKENHI Grant Number 26400414 and by MEXT
KAKENHI Grant Number 25103007.
\end{acknowledgments}


\begin{thebibliography}{99}
\bibitem{Note}
Although a Skyrmion ground state does not exist for $g_{11} = g_{22} =
g_{12}$~\cite{Battye}, there exists a stable stationary state traveling at
a constant velocity.
\end{thebibliography}
\end{document}